\documentclass[DIV=calc,paper=a4,fontsize=11pt,twocolumn]{scrartcl} 

\usepackage[english]{babel}
\usepackage[protrusion=true,expansion=true]{microtype}
\usepackage{amsmath,amsfonts,amsthm}
\usepackage[final]{graphicx}
\usepackage{xcolor}
\usepackage[normal,small,hypcap,up,labelfont=bf,textfont=it]{caption}
\usepackage{epstopdf}
\usepackage{subfig}
\usepackage{booktabs}
\usepackage{fix-cm}
\usepackage{amssymb,amsfonts}
\usepackage{dsfont}
\usepackage{bbm}
\usepackage{pstricks}
\usepackage{cite}
\usepackage[utf8]{inputenc}
\usepackage[perpage,symbol*]{footmisc}
\usepackage[varg]{txfonts}
\usepackage{balance}
\usepackage{fancyhdr}
\PassOptionsToPackage{hyphens}{url}\usepackage[pdfencoding=auto,psdextra]{hyperref}
\usepackage{bookmark}
\usepackage{verbatim}
\usepackage{fontenc}
\usepackage{cuted}

\usepackage{bm}
\usepackage{mathrsfs}

\theoremstyle{definition}

\theoremstyle{plain}

\DeclareCaptionFont{mycolor}{\color[HTML]{000000}}
\usepackage{sectsty}													
\allsectionsfont{
\color[HTML]{31ADF3}\usefont{OT1}{phv}{b}{n}
}

\sectionfont{
\color[HTML]{31ADF3}\usefont{OT1}{phv}{b}{n}
}

\usepackage{fancyhdr}												
\usepackage{lettrine}
\newcommand{\initial}[1]{%
\lettrine[lines=3,lhang=0.3,nindent=0em]{
\color[HTML]{31ADF3}
{\textsf{#1}}}{}}

\usepackage{titling}															

\newcommand{\HorRule}{\color[HTML]{31ADF3}
\rule{\linewidth}{1pt}%
}

\pretitle{\vspace{-30pt} \begin{flushleft} \HorRule
\fontsize{34}{34} \usefont{OT1}{phv}{b}{n} \color[HTML]{31ADF3} \selectfont
}
\title{Biphoton Interference in a Double-Slit Experiment}					
\posttitle{\par\end{flushleft}\vskip 0.5em}

\preauthor{\begin{flushleft}\large \lineskip 0.5em \usefont{OT1}{phv}{b}{sl} \color[HTML]{31ADF3}}
\author{Ananya Paul$^{~\mathsf{1}}$ \& Tabish Qureshi$^{~\mathsf{2}}$\\[8pt]}											
\postauthor{\footnotesize \usefont{OT1}{phv}{m}{sl} \color[HTML]{000000}
$^{\mathsf{1}}$ Department of Physics, Jamia Millia Islamia, New Delhi, India. E-mail: \href{mailto:ananya94ananya@gmail.com}{ananya94ananya@gmail.com}\\
$^{\mathsf{2}}$ Centre for Theoretical Physics, Jamia Millia Islamia, New Delhi,
India. E-mail: \href{mailto:tabish@ctp-jamia.res.in}{tabish@ctp-jamia.res.in}\\[10pt]		
\scriptsize\usefont{OT1}{phv}{m}{n} \color[HTML]{31ADF3}{\textbf{Editors: \emph{Chariton Aris Chatzidimitriou-Dreismann} \& \emph{Danko Georgiev}} }\\[5pt]
\color[HTML]{000000}{Article history: Submitted on October 30, 2017;  Accepted on February 19, 2018; Published on February 20, 2018.}
\par\end{flushleft}\HorRule}

\date{}																				

\begin{document}
\maketitle
\thispagestyle{fancy} 			
\initial{A}\textbf{double-slit experiment with entangled photons is theoretically analyzed.
It is shown that, under suitable conditions, two entangled photons of
wavelength $\lambda$ can behave like a \emph{biphoton} of wavelength $\lambda/2$.
The interference of these biphotons, passing through a double-slit can
be obtained by detecting both photons of the pair at the same position.
This is in agreement with the results of an earlier experiment.
More interestingly, we show that even if the two entangled photons are 
separated by a polarizing beam splitter, they can still behave like 
a biphoton of wavelength $\lambda/2$. In this modified setup, the two
separated photons passing through two different double-slits, surprisingly show
an interference corresponding to a wavelength $\lambda/2$, instead of
$\lambda$ which is the wavelength of each photon.
We point out two experiments that have been carried out in different
contexts, which saw the effect predicted here without realizing this connection.\\ Quanta 2018; 7: 1--6.}

\section{Introduction}

Quantum mechanics has taught us that wave nature and particle nature are two
complementary aspects of the same entity \cite{bohr}. Whether we talk of massive particles
or quanta of light, both can behave like particles and waves in different
situations. Young's double-slit experiment carried out with individual
particles showed that a particle passes through two slits and interferes
with itself \cite{jonsson}. Later it was demonstrated that much larger
particles such as $C_{60}$ molecules can also show interference
\cite{buckyball}.  It has been convincingly argued that instead of calling
them waves or particles, such entities should be
called \emph{quantons} \cite[p.~235]{bunge}\cite{levy}. Going beyond this, quantum mechanics
also tells us that a group of entities, e.g., many photons
together, can behave as a single quanton. Consequences of this on interference
experiments with many particles, has only been recognized
relatively recently \cite{multiphoton}.

First, we briefly explain the idea which motivated Jacobson and collaborators
\cite{multiphoton} to propose that many photons can behave as a single
quanton in an interference experiment. Consider a beam of  diatomic iodine molecules $I_2$
each with mass $2m$, traveling with a velocity $v$, passing through a
double-slit. The resulting interference would be in
accordance with a de Broglie wavelength $\lambda_{2m} = h/2mv$. But suppose that
the molecule dissociates on the way, and only separate iodine atoms, each of
mass $m$, pass through the double-slit. Then the resulting interference would
be in accordance with a de Broglie wavelength $\lambda_m = h/mv$, which
shows that $\lambda_{2m} = \lambda_m/2$.
More generally, $N$ particles with a de Broglie wavelength $\lambda$, can
behave as single quanton of wavelength $\lambda/N$. The same should hold for
photons too. An experiment was subsequently carried out which measured the
de Broglie wavelength of a two-photon wavepacket \cite{multiphotonexpt}.

\begin{figure}[b!]
\rule{245 pt}{0.5 pt}\\[3pt]
\raisebox{-0.2\height}{\includegraphics[width=5mm]{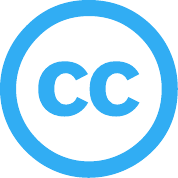}}\raisebox{-0.2\height}{\includegraphics[width=5mm]{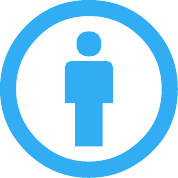}}
\footnotesize{This is an open access article distributed under the terms of the Creative Commons Attribution License \href{http://creativecommons.org/licenses/by/3.0/}{CC-BY-3.0}, which permits unrestricted use, distribution, and reproduction in any medium, provided the original author and source are credited.}
\end{figure}
\pagebreak
In the following we carry out a wave-packet analysis of two entangled photons,
typically generated in a type-I spontaneous parametric down conversion (SPDC)
process, and analyze the situation in which they can behave like a single
quanton.

\begin{figure}
\centerline{\resizebox{8.0cm}{!}{\includegraphics{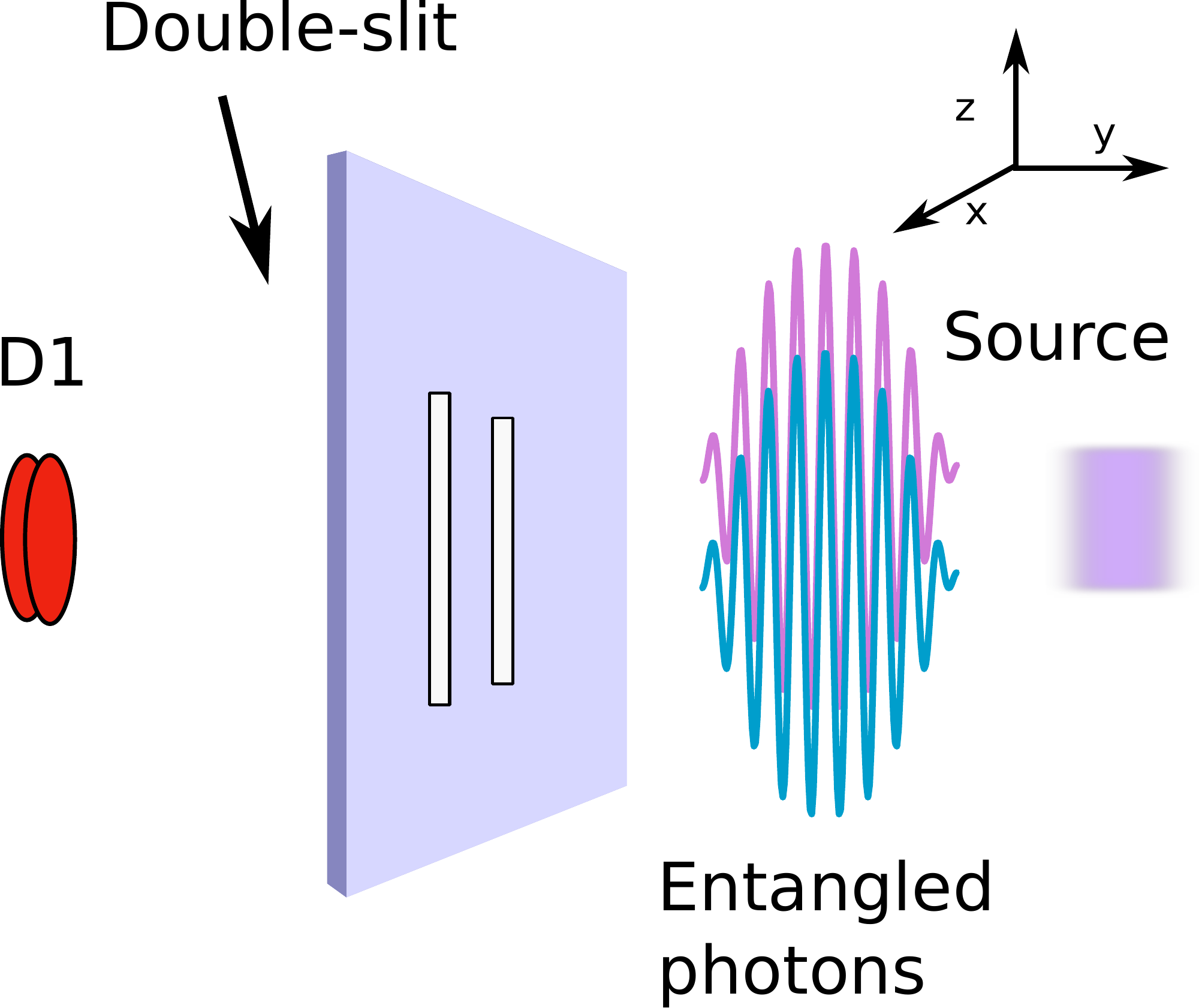}}}
\caption{Schematic diagram for Young's double-slit experiment with
entangled photons. Detector D1 is capable of detecting pairs of photons.
It should be able to discriminate between one-photon and two-photons events.
}
\label{2slit}
\end{figure}

\section{Theoretical analysis}
\subsection{Entangled photons}

A well known state to describe momentum-entangled particles was discussed by
Einstein, Podolsky and Rosen (EPR) \cite{epr}
\begin{equation}
\Psi_{\textrm{EPR}}(x_1,x_2) = A\!\int_{-\infty}^\infty 
e^{-{\imath px_2\over\hbar}} e^{\frac{\imath px_1}{\hbar}} dp. \label{epr}
\end{equation}
This so-called EPR state does capture the properties of entangled particles
well, but has some disadvantages like not being normalized, and also not
describing varying degree of entanglement.
The best state to describe momentum-entangled
particles is the \emph{generalized EPR state} \cite{tqajp,Qureshi2012}
\begin{equation}
\Psi(x_1,x_2) = A\!\int_{-\infty}^\infty 
e^{-{p^2\sigma^2\over \hbar^2}}e^{-{\imath px_2\over\hbar}} e^{\frac{\imath  px_1}{\hbar}}
e^{-{(x_1+x_2)^2\over 4\Omega^2}} dp, \label{state}
\end{equation}
where $A$ is a normalization constant, and $\sigma,\Omega$ are certain
parameters. In the limit $\sigma\to 0,~~\Omega\to\infty$ the state (\ref{state})
reduces to the EPR state (\ref{epr}).

After performing the integration over $p$, (\ref{state}) reduces to
\begin{equation}
\Psi(x_1,x_2) = {1\over \sqrt{\pi\sigma\Omega}}
 e^{\frac{-(x_1-x_2)^2}{4\sigma^2}} e^{\frac{-(x_1+x_2)^2}{4\Omega^2}} .
\label{gepr}
\end{equation}
It is straightforward to show that $\Omega$ and $\hbar/\sigma$ quantify the position
and momentum spread of the particles in the $x$-direction because the
uncertainty in position and the wave-vector of the two photons,
along the $z$-axis, is given by
\begin{equation}
\Delta x_1 = \Delta x_2 = \sqrt{\Omega^2+\sigma^2},
\Delta k_{1x} = \Delta k_{2x} = 
\tfrac{1}{4}\sqrt{{1\over \sigma^2} + {1\over \Omega^2}}~. \label{unc}
\end{equation}
Notice that for
$\sigma = \Omega$, the state is no longer entangled, and factors into
a product of two Gaussians centered at $x_1=0$ and $x_2=0$, respectively.
The state (\ref{gepr}) also describes well the two-photon mode function
at the output of the type-I crystal in SPDC generation \cite{photons1,photons2}.

The experiment is schematically described in Figure~\ref{2slit}. Entangled
particles (generally photons) emerge from a source, and pass through
a double-slit to reach a screen or a detector D1 which is movable along the
$x$-axis.
We assume that at time $t=0$, the two particles are in the state (\ref{gepr}),
and travel along the $y$-axis, towards a double-slit, with average momenta
$p_0$. Each particle can then be described as a quanton with a
wavelength $\lambda=h/p_0$. For photons, the wavelength is fixed as
$\lambda=2\pi/k_0$. 

\subsection{Time evolution}

Time evolution of a one-dimensional wave-packet, along $x$-axis, is given by
\begin{equation}
\psi(x,t) = {1\over\sqrt{2\pi}}\int_{-\infty}^{\infty} \tilde{\psi}(k_x)
\exp\left[\imath (k_xx-\omega(k_x)t)\right]dk_x.
\end{equation}
For massive particles, one would have assumed $\omega(k_x)=\hbar k_x^2/2m$.
For photons one can work within the Fresnel approximation, ($k_y\approx k_0$,
$k_x \ll k_y$) to write $\omega(k_x)$ as \cite{dillon}
\begin{equation}
\omega(k_x) = c\sqrt{k_x^2+k_y^2} \approx ck_0 +  {ck_x^2\over 2k_0}.
\end{equation}
So the spread of a photon wave-packet in the $x$-direction, which is moving
essentially along $y$-direction, is given by
\begin{equation}
\psi(x,t) = {e^{-\imath k_0t}\over\sqrt{2\pi}}\int_{-\infty}^{\infty} \tilde{\psi}(k_x)
e^{\imath k_xx}e^{-\imath ctk_x^2/2k_0} dk_x~.
\end{equation}

Using the above, the time
propagation kernel for the two photons can be written as
\begin{eqnarray}
K_1(x_1,x_1',t) &=& \sqrt{1\over \imath \lambda ct}
\exp\left[{-\pi(x_1-x_1')^2\over \imath \lambda ct}\right],\nonumber\\
K_2(x_2,x_2',t) &=& \sqrt{1\over \imath \lambda ct}
\exp\left[{-\pi(x_2-x_2')^2\over \imath \lambda ct}\right],
\end{eqnarray}
and the two-particle state after a time $t$ is given by
\begin{eqnarray}
\Psi(x_1,x_2,t) &=& \int_{-\infty}^{\infty} \int_{-\infty}^{\infty}K_1(x_1,x_1',t)\times\nonumber\\
 && \qquad K_2(x_2,x_2',t)
\Psi(x_1',x_2')~ dx_1' dx_2'.\nonumber\\
\end{eqnarray}

At this stage it is convenient to introduce new coordinates for the 
entangled particles: $r=(x_1+x_2)/2,~~q=(x_1-x_2)/2$. The state of the
entangled particles, at time $t=0$, can then be written as
\begin{equation}
\Psi(r,q) = {1\over \sqrt{\pi\sigma\Omega}}
 e^{-q^2/\sigma^2} e^{-r^2/\Omega^2} .
\label{gepr1}
\end{equation}
The time-propagator, in the new coordinates, can be written as
\begin{eqnarray}
K_r(r,r',t) &=& \sqrt{1\over \imath \lambda ct}
\exp\left[{-2\pi(r-r')^2\over \imath \lambda ct}\right]\nonumber\\
K_q(q,q',t) &=& \sqrt{1\over \imath \lambda ct}
\exp\left[{-2\pi(q-q')^2\over \imath \lambda ct}\right].
\label{propagator}
\end{eqnarray}
The state after a general time $t$ can then be evaluated as
\begin{eqnarray}
\Psi(r,q,t) &=& \int_{-\infty}^{\infty}\int_{-\infty}^{\infty} K_r(r,r',t)\times\nonumber\\
&& \qquad K_q(q,q',t)
\Psi(r',q')~dr' dq'.
\end{eqnarray}
Let us assume that during a time $t_0$, the photons travel a distance $L$,
from the source to
the double-slit, and the state at the double-slit takes the form:
\begin{equation}
\Psi(r,q,t_0) = C \exp\left({-q^2\over\sigma^2+\imath \alpha}\right)
\exp\left({-r^2\over\Omega^2+\imath \alpha}\right) ,
\label{t_slit}
\end{equation}
where $C=\frac{1}{\sqrt{\pi\sqrt{\sigma+\imath {\alpha/\sigma}}
\sqrt{\Omega+\imath {\alpha/\Omega}}}}$, and $\alpha = \lambda L/2\pi$.

\subsection{Effect of the double-slit}

After a time $t_0$, the two photons reach the double-slit and pass through it
to emerge on the other side. A rigorous, but immensely difficult approach
would be to consider the double-slit as a potential, and let the two photons
evolve under the action of that potential. We take a simpler and less
rigorous approach, by assuming that the effect of the double-slit is to
truncate the wave-function abruptly such that only the part of the wave
function in the region $-{d\over 2}-{\epsilon\over 2} \le x_1,x_2 \le 
-{d\over 2}+{\epsilon\over 2}$ and
${d\over 2}-{\epsilon\over 2} \le x_1,x_2 \le 
{d\over 2}+{\epsilon\over 2}$ survives. This region corresponds to the
region of the two slits, if the slits of width $\epsilon$ are located
at $x=-{d\over 2}$ and $x={d\over 2}$.
In our new coordinates, this region corresponds approximately to
(a) $\pm{d\over 2}-{\epsilon\over 2} \le r \le 
\pm{d\over 2}+{\epsilon\over 2}$ together with
$-{\epsilon\over 2} \le q \le {\epsilon\over 2}$
and
(b) $\pm{d\over 2}-{\epsilon\over 2} \le q \le 
\pm{d\over 2}+{\epsilon\over 2}$ together with
$-{\epsilon\over 2} \le r \le {\epsilon\over 2}$.
This is not completely accurate as far as $\epsilon$ is concerned, but
since the interference will be seen in the limit of very small $\epsilon$, 
this approximation suffices for our purpose.
Case (a) corresponds to both photons passing through the same slit,
whereas case (b) corresponds to both photons passing through
different slits. Notice that if the two photons have a high spatial
correlation, case (b) is expected to have very low probability.

The two photons travel a distance $D = ct$ to reach the screen/detector.
The state at the screen is given by the following time-evolution
\begin{eqnarray}
\Psi(r,q,t) &=& \int_{-{d\over 2}-{\epsilon\over 2}}^{-{d\over 2}+{\epsilon\over 2}}dr'K_r(t)
\int_{-{\epsilon\over 2}}^{{\epsilon\over 2}}dq'K_q(t) \Psi(r',q',t_0)\nonumber\\
&+& \int_{{d\over 2}-{\epsilon\over 2}}^{{d\over 2}+{\epsilon\over 2}}dr'K_r(t)
\int_{-{\epsilon\over 2}}^{{\epsilon\over 2}}dq'K_q(t) \Psi(r',q',t_0)\nonumber\\
&+& \int_{-{d\over 2}-{\epsilon\over 2}}^{-{d\over 2}+{\epsilon\over 2}}dq'K_q(t)
\int_{-{\epsilon\over 2}}^{{\epsilon\over 2}}dr'K_r(t) \Psi(r',q',t_0)\nonumber\\
&+& \int_{{d\over 2}-{\epsilon\over 2}}^{{d\over 2}+{\epsilon\over 2}}dq'K_q(t)
\int_{-{\epsilon\over 2}}^{{\epsilon\over 2}}dr'K_r(t) \Psi(r',q',t_0) ,\nonumber\\
\label{psiformal}
\end{eqnarray}
where the propagator and the initial state are given by (\ref{propagator}) and
(\ref{t_slit}), respectively. For brevity, the $q,q',r,r'$ dependence of
the propagators has been suppressed.
A typical integral in (\ref{psiformal}) looks like the following:
\begin{eqnarray}
I &=& \int_{{d\over 2}-{\epsilon\over 2}}^{{d\over 2}+{\epsilon\over 2}}
\exp\left[{-{2\pi(r-r')^2\over \imath \lambda L}}\right]
\exp\left[{-{r'^2\over\Omega^2+\imath \alpha}}\right] dr'.\nonumber\\
\label{term}
\end{eqnarray}
Since the profile of the incoming beam is wide, $\Omega^2 \gg \lambda L/2\pi$.
The slit width $\epsilon$ is assumed to be very small. Since in the integral
above, $r'$ varies only between ${d\over 2}-{\epsilon\over 2}$ to
${d\over 2}+{\epsilon\over 2}$, the term
$\exp\left({-{r'^2\over\Omega^2+\imath \alpha}}\right)$
can be assumed to be constant in this region, and equal to
$\exp\left({-{d^2/4\over\Omega^2+\imath \alpha}}\right)$. Keeping in mind the
smallness of $\epsilon$, we can make an additional approximation,
$(r-r')^2 \approx (r-{d\over 2})^2-2(r-{d\over 2})(r'-{d\over 2})$,
ignoring terms of order $\epsilon^2$. With these assumptions, the
integral in (\ref{term}) can be approximated by
\begin{eqnarray}
I &\approx& e^{2\pi \imath (r-d/2)^2\over\lambda D}
e^{-{d^2/4\over\Omega^2+\imath \alpha}}
\int_{{d\over 2}-{\epsilon\over 2}}^{{d\over 2}+{\epsilon\over 2}}
e^{-{4\pi\imath (r-d/2)(r'-d/2)\over \lambda D}}dr' \nonumber\\
  &=& e^{2\pi \imath (r-d/2)^2\over\lambda D}
e^{-{d^2/4\over\Omega^2+\imath \alpha}}
\tfrac{\sin\left(2\pi(r-d/2)\epsilon/\lambda D\right)}{ 2\pi(r-d/2)/\lambda D}.
\end{eqnarray}
If similar algebra is carried out over all the integrals in (\ref{psiformal}),
one obtains the following form of the final state of the biphoton
\begin{eqnarray}
\Psi(r,q,t) &=& C_t\left(e^{{2\pi\imath \over\lambda D}(r-{d\over 2})^2}
e^{{2\pi\imath \over\lambda D}q^2}f(r-\tfrac{d}{2})f(q)
e^{-{d^2\Omega^2\over4\Omega^4+4\alpha^2}}\right.\nonumber\\
&&\left.+ e^{{2\pi\imath \over\lambda D}(r+\tfrac{d}{2})^2}
e^{{2\pi\imath \over\lambda D}q^2}f(r+\tfrac{d}{2})f(q)
e^{-{d^2\Omega^2\over4\Omega^4+4\alpha^2}}\right.\nonumber\\
&&\left.+ e^{{2\pi\imath \over\lambda D}(q+\tfrac{d}{2})^2}
e^{{2\pi\imath \over\lambda D}r^2}f(q+\tfrac{d}{2})f(r)
e^{-{d^2\sigma^2\over4\sigma^4+4\alpha^2}}\right.\nonumber\\
&&\left.+ e^{{2\pi\imath \over\lambda D}(q-\tfrac{d}{2})^2}
e^{{2\pi\imath \over\lambda D}r^2}f(q-\tfrac{d}{2})f(r)
e^{-{d^2\sigma^2\over4\sigma^4+4\alpha^2}}\right) ,\nonumber\\
\label{finalstate}
\end{eqnarray}

\begin{figure*}
\centerline{\resizebox{155mm}{!}{\includegraphics{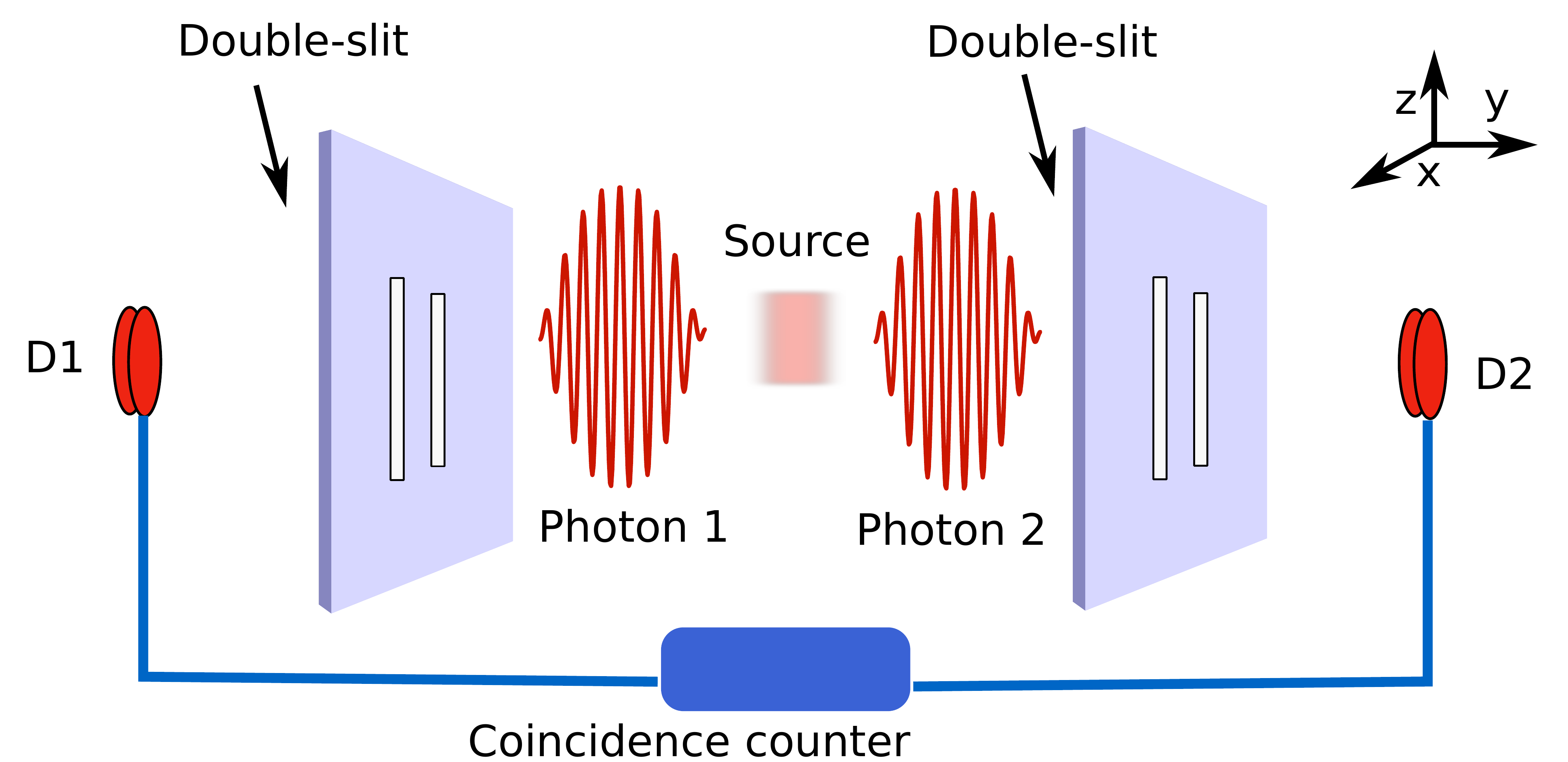}}}
\caption{Schematic diagram for the proposed nonlocal biphoton experiment.
Photons 1 and 2 effectively move in opposite directions along $y$-axis.
Detectors D1 and D2 move along the $x$-axis in synchrony such that their 
$x$-positions are always the same. They also count the photons in coincidence.
}
\label{nonlocal}
\end{figure*}

\noindent where $C_t={\imath \over\lambda D}(\pi)^{-1/2}(\sigma+\imath {\imath \alpha\over\sigma})^{-1/4}
(\Omega+\imath {\imath \alpha\over\Omega})^{-1/4}$, 
and $f(x)\equiv \frac{\sin\left({2\pi x\epsilon/\lambda D}\right)}
{2\pi x/\lambda D}$ governs the spatial spread of the interference pattern.
When the spatial spread of the biphoton at the 
double-slit is much larger than the slit separation, the term
$e^{-{d^2\Omega^2\over4\Omega^4+4\alpha^2}}$ is of the order of unity.
If the spatial correlation between the two photons is high at the
double-slit, $\sigma$ is very small and consequently, the term
$e^{-{d^2\sigma^2\over4\sigma^4+4\alpha^2}}$  becomes much smaller
than unity. For $\epsilon \ll 1$, in a large region around $r=0$ on the
screen, we can make
the approximation $f(r-\tfrac{d}{2})\approx f(r+\tfrac{d}{2}) \approx f(r)$.
One may note that because of the truncation approximation, the state
(\ref{finalstate}) is no longer normalized. However, since we are only
interested in the interference pattern, we will continue to work with the
unnormalized state.

\section{Results}

\subsection{Biphoton with wavelength $\lambda/2$}

If the entanglement between the two photons is good, the last two terms
in (\ref{finalstate}) can be dropped. One would like to see the
distribution of the two photons striking at the same position on the screen.
This can be achieved by putting $x=(x_1+x_2)/2=r$ and $q = (x_1-x_2)/2=0$.
The probability density $P(x)$ of the two photons striking together at a position
$x$ on the screen is then given by $|\Psi(x,0,t)|^2$ where $\Psi$ is given
by (\ref{finalstate}). Within the approximations described above, the
probability density of the biphoton to strike a position $x$ on the screen
is given by
\begin{eqnarray}
P(x) &=& |C_t|^2\epsilon^2f^2(x)\left[1 + \cos\left({4\pi xd\over\lambda D}\right)\right].
\label{double}
\end{eqnarray}
The above expression represents an interference pattern with a fringe width
given by $w = {(\lambda/2)D\over d}$, which means that the biphoton behaves
like one quanton of wavelength $\lambda/2$.

This feature has already been experimentally demonstrated in an experiment
carried out with entangled photons generated via SPDC \cite{multiphotonexpt}.

\subsection{Nonlocal biphoton with wavelength $\lambda/2$}

We now argue that in order for the entangled photons to act as a single
quanton of wavelength $\lambda/2$, it is not necessary that they be
physically close together. That may sound like an outlandish claim, but we
shall see in the following how it may be possible.
We propose a modified experiment in which entangled
photons are separated by a polarizing beam-splitter, and each passes through
a different double-slit kept at equal distance from the beam splitter. 
Effectively, the photons may now be assumed to be traveling in opposite
directions along $y$-axis, as shown in Figure~\ref{nonlocal}. 

The two entangled photons, emerging from the source, are described by the
state (\ref{gepr}). They travel in opposite direction for a time $t_0$,
after which they reach their respective double-slits. The double-slits are kept
on opposite sides of the source, at a distance $L = ct_0$ from
the source. When the two photons reach the double-slits, their $x$-dependence
is described by (\ref{t_slit}). Of course, the $y$-dependence of the two
particles will be very different: one photon will be a wave-packet centered
at $y=-L$, and the other centered at $y=L$, assuming that the source sits at
$y=0$. However, as far as the
entanglement, and the \mbox{$x$-dependence} of the state is concerned, their
$y$-dependence is unimportant. We assume that the effect of the two double-slits
is to truncate the state of the two photons to the region within the slits,
i.e.,
${-{d\over 2}}-{\epsilon\over 2} \le x_1,x_2 \le 
{-{d\over 2}}+{\epsilon\over 2}$ and
${d\over 2}-{\epsilon\over 2} \le x_1,x_2 \le 
{d\over 2}+{\epsilon\over 2}$.
Needless to say that for this argument to work, the $x$-positions of the two
double-slits should be exactly the same. This would make sure that the two
photons, although traveling in different directions along $y$-axis, encounter
a slit at the same $x$-position, although their $y$-positions are separated.
It should be recalled that the two photons have a directional uncertainty
along the $x$-axis.
After emerging from the double-slits, the
two photons travel, for a time $t$, a distance $D = ct$, to reach their
respective detectors D1 and D2. The final state of the two photons at the
two detectors is given by (\ref{finalstate}). One would notice that the
same analysis, that was used for both photons traveling in the same direction
and passing through the same double-slit, works for the photons traveling in
opposite direction, and passing through different double-slits.

\begin{figure}
\centerline{\resizebox{8.5cm}{!}{\includegraphics{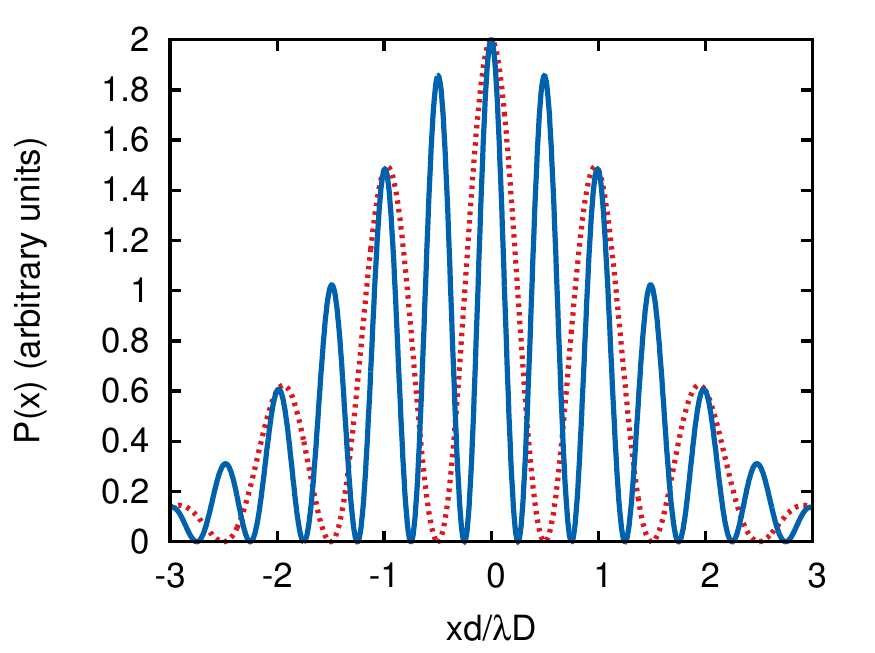}}}
\caption{Double-slit interference pattern of the biphoton given by
(\ref{double}), where $\lambda$ is the wavelength of the photons (solid line).
Fringe width is $w = {(\lambda/2) D\over d}$.
Double-slit interference pattern of the photons given by (\ref{single})
(dotted line). Fringe width is $w = {\lambda D\over d}$. A typical profile
of $f(x)$ has been used for the plots.
}
\label{interf}
\end{figure}

The probability density of coincident click of D1 at $x_1 = x$ and 
D2 at $x_2 = x$, is given by
$P(x) = |C_t|^2\epsilon^2f^2(x)\left[1 + \cos\left({4\pi xd\over\lambda D}\right)\right]$, which is the same as (\ref{double}).
But this is an interference pattern corresponding to a wavelength $\lambda/2$.
Thus we reach an amazing conclusion, that the two photons, although widely
separated in space, behave like a single quanton of wavelength $\lambda/2$
which interferes with itself (see Figure \ref{interf}).

Interestingly, an experiment with entangled photons was carried out in the
context of quantum lithography, which showed the effect predicted here,
namely, the interference pattern appearing corresponding to a wavelength
$\lambda/2$, where $\lambda$ is the wavelength of the photons \cite{shihlitho}.
However, the authors of the experiment have not analyzed it in the light of
multiphoton deBroglie waves  \cite{multiphoton,multiphotonexpt}.

Another experiment with electrons emitted from photodouble ionization
of $H_2$ molecules has been performed very recently, which seems to show
an effect closely related to the one predicted here \cite{dielectron}.
The two electrons do not pass through any double-slit, but are produced at
two indistinguishable centers A or B separated by the internuclear distance
of two atoms in the hydrogen molecule.
The authors concluded that the two electrons behave like a \emph{dielectron}
which has a wave-vector of magnitude $k_1+k_2$, $k_1, k_2$ being the
magnitudes of the wave-vectors of the two electrons. It is easy to see that
had the two wave-vectors been of the same magnitude, the \emph{dielectron}
would have a de Broglie wavelength half the wavelength of a single electron.
The authors of this paper too, have not connected their results to
the earlier work on multiphoton interference \cite{multiphoton,multiphotonexpt}.

\subsection{Single photon interference}

We now investigate the possibility of a photon of the entangled pair behaving
like a standalone quanton. This can be achieved by fixing detector D2 at
$x_2=0$ and counting photons by D1 at various $x_1$, in coincidence with D2.
Putting $x_2=0$ corresponds to $r = x_1/2$ and $q = x_1/2$. Doing that
simplifies (\ref{finalstate}) to
\begin{eqnarray}
\Psi(x_1,x_2=0,t) &\approx& C_t\left(e^{{\pi\imath \over 2\lambda D}(x_1-d)^2}
e^{{\pi\imath \over 2\lambda D}x_1^2}f^2(\tfrac{x_1}{2})
\right.\nonumber\\
&&\left.+ e^{{\pi\imath \over 2\lambda D}(x_1+d)^2}
e^{{\pi\imath \over 2\lambda D}x_1^2}f^2(\tfrac{x_1}{2})
\right),
\end{eqnarray}
where the combined state $\Psi(x_1,x_2,t)$ is labeled by the original 
coordinates $x_1,x_2$, and not by $r, q$. The probability density to find
a photon at $x_1$, $P(x_1)$ is given by $P(x_1)=|\Psi(x_1,x_2=0,t)|^2$,
and has the following form:
\begin{eqnarray}
P(x_1) &=& |C_t|^2\epsilon^2f^2(\tfrac{x_1}{2})\left[1 + \cos\left({2\pi x_1d\over\lambda D}\right)\right].
\label{single}
\end{eqnarray}
The above represents a Young's double-slit interference pattern with a
fringe with $w = {\lambda D\over d}$. In this arrangement the photons
detected by D1 behave as independent quantons with wavelength $\lambda$
(see Figure \ref{interf}).

\section{Conclusions}

We have a done a wave-packet analysis of two entangled photons
passing through a double-slit. We have shown that the two photons can
behave like a single quanton of half the wavelength of the photons
when detected in coincidence at the same position. This is in agreement of
an earlier analysis and experiment by Fonseca, Monken and P\'adua
\cite{multiphotonexpt}. Going further, we have shown that the two photons can
continue to behave like a single quanton even when they are widely separated in space, a
highly nonlocal feature. This work extends the theoretical ideas of
multiphoton wave packets \cite{multiphoton,multiphotonexpt} to a nonlocal
scenario. Our result implies that even when two entangled photons are
separated in space, they may act like a single quanton which interferes 
with itself. Entangled particles show very strange and counter-intuitive
properties. It has previously been shown that entangled photons can
exhibit a \emph{nonlocal} wave-particle duality \cite{gduality}.

\section*{Acknowledgment}
Ananya Paul is thankful to the Centre for Theoretical Physics, Jamia Millia
Islamia, New Delhi, for providing its facilities during the course of this
work.


\begin{thebibliography}{10}
\balance
\bibitem{bohr}
Bohr N.
The quantum postulate and the recent development of atomic theory.
\emph{Nature} 1928; \textbf{121}(3050): 580--590. \href{http://dx.doi.org/10.1038/121580a0}{\path{doi:10.1038/121580a0}}

\bibitem{jonsson}
J\"{o}nsson C.
Electron diffraction at multiple slits.
\emph{American Journal of Physics} 1974; \textbf{42}(1): 4--11. \href{http://dx.doi.org/10.1119/1.1987592}{\path{doi:10.1119/1.1987592}} 

\bibitem{buckyball}
Nairz O, Arndt M, Zeilinger A.
Quantum interference experiments with large molecules.
\emph{American Journal of Physics} 2003; \textbf{71}(4): 319--325. \href{http://dx.doi.org/10.1119/1.1531580}{\path{doi:10.1119/1.1531580}} 


\bibitem{bunge}
Bunge M.
\emph{Foundations of Physics}. Springer Tracts in Natural Philosophy, vol.~10, New York: Springer, 1967. \href{http://dx.doi.org/10.1007/978-3-642-49287-7}{\path{doi:10.1007/978-3-642-49287-7}} 

\bibitem{levy}
L\'{e}vy-Leblond J-M.
Quantum words for a quantum world. In: \emph{Epistemological and Experimental Perspectives on Quantum Physics}. Greenberger D, Reiter WL, Zeilinger A (editors), Vienna Circle Institute Yearbook, vol.~7, Dordrecht: Springer, 1999, pp.~75--87. \href{http://dx.doi.org/10.1007/978-94-017-1454-9_5}{\path{doi:10.1007/978-94-017-1454-9_5}}

\bibitem{multiphoton}
Jacobson J, Bj\"{o}rk G, Chuang I, Yamamoto Y.
Photonic de Broglie waves.
\emph{Physical Review Letters} 1995; \textbf{74}(24): 4835--4838. \href{http://dx.doi.org/10.1103/PhysRevLett.74.4835}{\path{doi:10.1103/PhysRevLett.74.4835}} 

\bibitem{multiphotonexpt}
Fonseca EJS, Monken CH, P\'{a}dua S.
Measurement of the de Broglie wavelength of a multiphoton wave packet.
\emph{Physical Review Letters} 1999; \textbf{82}(14): 2868--2871. \href{http://dx.doi.org/10.1103/PhysRevLett.82.2868}{\path{doi:10.1103/PhysRevLett.82.2868}} 

\bibitem{epr}
Einstein A, Podolsky B, Rosen N.
Can quantum-mechanical description of physical reality be considered complete?
\emph{Physical Review} 1935; \textbf{47}(10): 777--780. \href{http://dx.doi.org/10.1103/PhysRev.47.777}{\path{doi:10.1103/PhysRev.47.777}} 

\bibitem{tqajp}
Qureshi T.
Understanding Popper's experiment.
\emph{American Journal of Physics} 2005; \textbf{73}(6): 541--544. \href{http://arxiv.org/abs/quant-ph/0405057}{\path{arXiv:quant-ph/0405057}}, \href{http://dx.doi.org/10.1119/1.1866098}{\path{doi:10.1119/1.1866098}}

\bibitem{Qureshi2012}
Qureshi T. Popper's experiment: a modern perspective. \emph{Quanta} 2012; \textbf{1}(1): 19--32. \href{http://dx.doi.org/10.12743/quanta.v1i1.8}{\path{doi:10.12743/quanta.v1i1.8}} 

\bibitem{photons1}
Chan KW, Torres JP, Eberly JH.
Transverse entanglement migration in Hilbert space.
\emph{Physical Review A} 2007; \textbf{75}(5): 050101. \href{http://arxiv.org/abs/quant-ph/0608163}{\path{arXiv:quant-ph/0608163}}, \href{http://dx.doi.org/10.1103/PhysRevA.75.050101}{\path{doi:10.1103/PhysRevA.75.050101}} 

\bibitem{photons2}
Di Lorenzo Pires H, van Exter MP.
Near-field correlations in the two-photon field.
\emph{Physical Review A} 2009; \textbf{80}(5): 053820. \href{http://dx.doi.org/10.1103/PhysRevA.80.053820}{\path{doi:10.1103/PhysRevA.80.053820}} 

\bibitem{dillon}
Dillon G.
Fourier optics and time evolution of de Broglie wave packets.
\emph{European Physical Journal Plus} 2012; \textbf{127}(6): 66. \href{http://arxiv.org/abs/1112.1242}{\path{arXiv:1112.1242}}, \href{http://dx.doi.org/10.1140/epjp/i2012-12066-2}{\path{doi:10.1140/epjp/i2012-12066-2}}  

\bibitem{shihlitho}
D'Angelo M, Chekhova MV, Shih Y.
Two-photon diffraction and quantum lithography.
\emph{Physical Review Letters} 2001; \textbf{87}(1): 013602. \href{http://arxiv.org/abs/quant-ph/0103035}{\path{arXiv:quant-ph/0103035}}, \href{http://dx.doi.org/10.1103/PhysRevLett.87.013602}{\path{doi:10.1103/PhysRevLett.87.013602}}  

\bibitem{dielectron}
Waitz M, Metz D, Lower J, Schober C, Keiling M, Pitzer M, Mertens K, Martins M, Viefhaus J, Klumpp S, Weber T, Schmidt-Böcking H, Schmidt LPH, Morales F, Miyabe S, Rescigno TN, McCurdy CW, Martín F, Williams JB, Schöffler MS, Jahnke T, Dörner R.
Two-particle interference of electron pairs on a molecular level.
\emph{Physical Review Letters} 2016; \textbf{117}(8): 083002. \href{http://arxiv.org/abs/1607.07275}{\path{arXiv:1607.07275}}, \href{http://dx.doi.org/10.1103/PhysRevLett.117.083002}{\path{doi:10.1103/PhysRevLett.117.083002}} 

\bibitem{gduality}
Siddiqui MA, Qureshi T.
A nonlocal wave-particle duality.
\emph{Quantum Studies: Mathematics and Foundations} 2016; \textbf{3}(1): 115--122. \href{http://arxiv.org/abs/1406.1682}{\path{arXiv:1406.1682}}, \href{http://dx.doi.org/10.1007/s40509-015-0064-4}{\path{doi:10.1007/s40509-015-0064-4}} 

\end{thebibliography}
\end{document}